\documentclass[twocolumn,showpacs,preprintnumbers,superscriptaddress]{revtex4}

\usepackage{amsfonts,amssymb,amsmath,amsthm}
\usepackage{graphicx}% Include figure files

\newtheorem{theorem}{Theorem}

\newtheorem{lemma}{Lemma}
\newtheorem{dfn}{Definition}

\newcommand{\calE}{{\cal E }}

\newcommand{\calS}{{\cal S }}

\newcommand{\la}{\langle}
\newcommand{\ra}{\rangle}

\newcommand{\be}{\begin{equation}}
\newcommand{\ee}{\end{equation}}
\newcommand{\nn}{\nonumber}
\newcommand{\ba}{\begin{array}}
\newcommand{\ea}{\end{array}}

\newcommand{\tr}{\mathop{\mathrm{Tr}}}

\newcommand{\aop}{{\hat{a}}}

\newcommand{\cop}{{\hat{c}}}

\newcommand{\xx}{{\bf x}}

\newcommand{\bdown}{b^{\downarrow}}

\newcommand{\ldown}{\lambda^{\downarrow}}
\newcommand{\ydown}{y^{\downarrow}}
\newcommand{\zdown}{z^{\downarrow}}
\newcommand{\pdown}{p^{\downarrow}}
\newcommand{\qdown}{q^{\downarrow}}

\begin{document}

\title{Classical capacity of fermionic product channels}

\author{Sergey Bravyi}
\email{serg@cs.caltech.edu}
\affiliation{Institute for Quantum Information,
California Institute of Technology,\\
Pasadena, 91125 CA, USA.}

\date{\today}

\begin{abstract}
We study multi-qubit quantum channels that can
be represented as a product of one-mode
fermionic attenuation channels.
An explicit formula for the classical capacity $C_1$
and for the minimum output
entropy $S_{min}$ of these channels is proposed.
We compute $S_{min}$ analytically for any number of
qubits under assumption that the minimum is achieved on
a Gaussian input. Apart from that, a simple numerical
method for evaluating $S_{min}$ is developed.
The method is applicable to any channels that are
sufficiently noisy. For fermionic product channels
the proposed formula for $S_{min}$ agrees with
the numerical results with a precision about $10^{-9}$.
\end{abstract}

\maketitle

\section{Introduction and summary of results}
Transmission of a classical information through a quantum communication
channel is one of the basic problems studied in the quantum
information theory. 
Suppose a sender (Alice) wants to send
a string of $n$ classical bits through a one-way noisy quantum channel
$\Phi$ to a receiver (Bob).
Suppose Alice is allowed to use
the channel only $m$ times. A number $R$ is called an 
achievable rate if for any $\epsilon,\delta>0$ there exists a
transmission protocol with $n/m> R-\delta$ 
and an error probability smaller than $\epsilon$.
The classical capacity  of $\Phi$ 
is defined as the minimum number $C$
such that $R \le C$ for any achievable rate $R$.
A closely related quantity is a {\it one-shot capacity}
$C_1$ which is defined analogously to $C$ with a
restriction that Alice uses only 
{\it product states} for signaling. 
It has been shown that
$C(\Phi)=\lim_{m\to \infty}
C_1(\Phi^{\otimes m})/m$ and conjectured that
actually $C(\Phi)=C_1(\Phi)$, i.e.,  $C_1$ is 
additive under tensor product of channels.
The Holevo-Schumacher-Westmoreland theorem~\cite{SW97,Holevo98}
states that
\be\label{C_1}
C_1(\Phi)=\sup_{\calE} \chi(\calE)
\ee
where $\calE=\{p_a,\rho_a\in Im(\Phi)\}$
is a probabilistic ensemble of states 
from the image of $\Phi$ (a set of all output states),
and $\chi(\calE)$ is the Holevo quantity
\be\label{Holevo}
\chi(\calE)=S\left(\sum_a p_a \rho_a\right)-\sum_a p_a \, S(\rho_a).
\ee

Important examples of quantum channels for which an 
explicit formula for the classical capacity is known
are product of one-qubit unital channels~\cite{King02}
and products of bosonic attenuation channels~\cite{MIT04}.
In the latter case a restriction on a power of
the input signal has to be imposed to regularize
the capacity. An explicit formula for the one-shot
capacity is also conjectured for bosonic attenuation
channels combined with a classical Gaussian noise~\cite{HW01}.

Quantum channels explored in this paper are direct
fermionic analogues of bosonic product channels
studied in~\cite{MIT04}. In contrast to their bosonic
counterparts, fermionic modes are described by a 
finite-dimensional Hilbert space, so one does not
need to regularize the capacity. 

An algebra of observables of $n$ fermionic modes formally
coincides with the one of $n$ qubits and can be conveniently
described by generators $\cop_1,\cop_2,\ldots,\cop_{2n}$,
\begin{eqnarray}\label{JW}
\cop_{2j-1} &=& \sigma^z_1\otimes \cdots \otimes \sigma^z_{j-1}\otimes
\sigma^x_j
\otimes I \otimes \cdots \otimes I, \nn \\
\cop_{2j} &=& \sigma^z_1\otimes \cdots \otimes \sigma^z_{j-1}\otimes
\sigma^y_j
\otimes I \otimes \cdots \otimes I. 
\end{eqnarray}
Here $\sigma^\alpha_j$ are the Pauli operators on the qubit $j$.
This is the well-known Jordan-Wigner transformation that is often used to
map a system of spins (qubits) into a system of fermions.
The generators $\cop_1,\ldots,\cop_{2n}$ obey the Clifford algebra 
commutation rules
\[
\cop_p\, \cop_q + \cop_q\, \cop_p = 2\, \delta_{pq} I, \quad
\cop_p^\dag=\cop_p.
\]
An arbitrary linear operator acting on $n$ qubits can be uniquely represented
as a linear combination of $4^n$ monomials
\[
\cop({\bf x})=\cop_1^{x_1}\, \cop_2^{x_2}\cdots \cop_{2n}^{x_{2n}},
\]
where ${\bf x}=(x_1,\ldots,x_{2n})$ is a binary string of $2n$ bits.

Define a quantum channel $\Phi$ by the following rules
\begin{eqnarray}
\label{Phi}
\Phi(I)&=&I,\nn \\
\Phi(\cop_p) &=& b_p\, \cop_p, \quad p=1,\ldots, 2n, \nn \\
\Phi(\cop({\bf x})) &=& \prod_{p\, : \, x_p=1} b_p\, 
 \cop({\bf x}). 
\end{eqnarray} 
Here $0\le b_1,\ldots,b_{2n}\le 1$ are 
$2n$ 
real parameters
specifying the channel (attenuation coefficients).
 A proof that
the linear map defined in
Eq.~(\ref{Phi}) is indeed
a quantum channel (i.e. a trace preserving completely
positive map) as well as some
motivation of this definition will be given in Section~\ref{sec:channel}.
It suffices to mention now that $\Phi$ has a clear product structure.
One can say that the system of $n$ qubits is `partitioned' into $2n$
separated subsystems (Majorana fermionic modes) that are described by
the operators $\cop_1,\ldots,\cop_{2n}$. Each subsystem is transmitted
independently through its own `wire' described by a quantum
channel $\cop_p \to b_p \, \cop_p$. 
However this product structure
should not be mixed with the tensor product of quantum channels.
Indeed, we shall see in Section~\ref{sec:channel} that Krauss
operators corresponding to different `wires' do not commute.
Accordingly, one can not expect that the capacity of the
channel Eq.~(\ref{Phi}) will be additive
(a capacity of an individual wire is not even a
well-defined quantity).

To compute the maximum at Eq.~(\ref{C_1}) we shall use
a certain variational family of states, namely, fermionic
Gaussian states~\cite{TD02,Bravyi05}.
A Gaussian state $\rho$ is completely
specified by its first and second moments $\tr(\rho\, \cop_p)$,
$\tr(\rho\, \cop_p \, \cop_q)$, see Section~\ref{sec:gauss}
for a strict definition.
In particular, the von Neumann
entropy $S(\rho)$ is a simple function of these moments.
We will see that the product
fermionic channels map Gaussian states into Gaussian states.
Define a {\it Gaussian capacity} as
\be\label{C_gauss}
C_1^g(\Phi)
=\sup_{\calE_g} \chi(\calE_g),
\ee
where $\calE_g=\{p_a,\rho_a\in Im(\Phi)\}$ is an ensemble
of Gaussian states (probabilities $p_a$ may be arbitrary 
though).
This  is the best transmission rate that can be achieved if
Alice uses only product Gaussian states for signalling. 
Obviously, $C_1^g\le C_1$. Our main result is an
explicit formula for the Gaussian capacity.

\begin{theorem}\label{thm:main}
Let $\Phi$ be a fermionic product channel
acting on $n$ qubits
with attenuation coefficients $0\le b_1,\ldots,b_{2n}\le 1$.
Denote $\bdown_j$ the $j$-largest coefficient.
Then 
\be\label{main_result}
C_1^g(\Phi)=n-H\left(\frac{1+ \bdown_1}2\right) 
-
\sum_{j=1}^{n-1} H\left(\frac{1+ \bdown_{2j}\, \bdown_{2j+1}}2\right),
\ee
where $H(x)=-x\log(x)-(1-x)\log (1-x)$ is the Shannon binary entropy.
Moreover, $C_1(\Phi)=C_1^g(\Phi)$ if $n\le 2$.
\end{theorem}
A proof of the theorem as well as some additional results
that are valid for Gaussian input states is given in
Section~\ref{sec:minimum}.

The theorem implies that the optimal signaling ensemble
is Gaussian for $n\le 2$. Can one achieve a higher 
transmission rate for $n>2$ by using non-Gaussian ensembles?
To get some intuition about it, we 
have tried to calculate $C_1$ numerically 
for three and four qubits, see Section~\ref{sec:numerics}.
The numerical results 
strongly suggest that 
$C_1=C_1^g$, i.e., that we gain nothing from using
non-Gaussian signaling states.
It might seem rather surprising because mixed Gaussian states are
known to be the maximally noisy states (in terms of their
von Neumann entropy) for fixed first and second moments. 
On the other hand, one can easily see from Eq.~(\ref{Phi})
that the channel $\Phi$ washes out
correlations among large number of modes such that
$l$-mode correlators acquire a factor $b^l$, where $b$ is a typical
value of $b_p$. It means that two-mode correlations is
the best place to keep the information that has to be sent
through the channel.
In this respect Gaussian states are very promising candidates,
because they have {\it only} two-mode correlations. 
Therefore, there might be a subtle tradeoff between large
entropy of Gaussian states and their special correlation structure
that makes them optimal for signaling.

In general, a computation of $C_1$ (even numerical) is an
extremely hard problem. Fortunately, the fermionic product channels
possess a special symmetry property known
as {\it covariance}, see~\cite{Holevo02,Cortese02}.
For covariant $n$-qubit channels
the one-shot capacity is given by 
\be\label{covariant}
C_1(\Phi)=n-S_{min}(\Phi),
\ee
where 
\be\label{Smin}
S_{min}(\Phi)=\min_{\rho} S(\Phi(\rho))
\ee
is the minimum output entropy. Our derivation of the Gaussian
capacity is actually a computation of the minimum output
entropy for Gaussian input states. 
The analogous problem has been already solved for
bosonic Gaussian channels~\cite{MIT03} and a 
conjecture has been made that the global minimum
of the output entropy is achieved on a Gaussian input.
Whether or not this conjecture is true in the fermionic world
is an open question
(which is equivalent to an equality $C_1=C_1^g$).

In Section~\ref{sec:numerics} we describe a simple
algorithm that allows one to evaluate $S_{min}(\Phi)$
for many quantum channels. Applicability of the algorithm is
not restricted to the fermionic product channels.
In general, the algorithm 
works well if $\Phi$ is not too close to an ideal channel.
The main idea is to exploit positivity of the relative 
entropy and iteratively minimize $S(\Phi(|\psi\ra\la\psi|))$.
More specifically, 
we construct a sequence of pure states
$|\psi_0\ra,|\psi_1\ra,|\psi_2\ra,\ldots$, such that
$|\psi_{j+1}\ra$ is the largest eigenvector
of an operator $\Phi^*(\log \Phi(|\psi_j\ra\la \psi_j|))$.
The initial state $|\psi_1\ra$ is chosen randomly.
One can easily show that 
$S(\Phi(\psi_{j+1}))\le S(\Phi(\psi_j))$,
i.e., the corresponding sequence of output entropies
is non-increasing. One should repeat the iterations
until the output entropy keeps decreasing.

In contrast to local search methods, such as
the gradient descent, the iterative minimization is able
to escape from local minimums of the objective
function. However, one can not guarantee that
the iterations converge to the global minimum,
see Section~\ref{sec:numerics} for more detailed
discussion. 

In Section~\ref{sec:numerics}.B we use the algorithm
to find the minimum output entropy of the fermionic product channels
for three-qubit and four-qubit systems.
The corresponding value of the capacity $C_1$
matches the Gaussian capacity $C_1^g$ with  a precision
about $10^{-9}$ which has the same order of magnitude as
the numerical noise.  It is therefore fair to make a conjecture
that $C_1=C_1^g$ in general.

The rest of the paper is organized as follows.
Section~\ref{sec:channel} contains all the necessary
facts about fermionic product channels.
In Section~\ref{sec:gauss} we review some basic 
properties of Gaussian states. Computation
of the minimal output entropy for Gaussian input 
is done in Section~\ref{sec:minimum}. 
Also in this section we show that the image
of a fermionic product channel has
a nicely ordered structure with respect to 
the majorization relation. The algorithm
for computation of the minimal output entropy 
and the numerical results obtained with its help
are discussed in Section~\ref{sec:numerics}.

\section{Basic properties of the fermionic product channel}
\label{sec:channel}

%We shall refer to the operators $\cop(x)$ as $\cop$-operators. 
%In other words, $\Phi$ is diagonal in the basis of $\cop$-operators.

\subsection{Krauss representation}
We start from constructing an explicit Krauss representation for
the channel $\Phi$ defined in Eq.~(\ref{Phi}).
Consider a {\it parity generator}
\be\label{parity}
P=(-i)^n\cop_1\cop_2\cdots\cop_{2n-1}\cop_{2n}.
\ee
In terms of qubits it looks as $P=\sigma^z_1\otimes \cdots
\otimes \sigma^z_n$.
One can easily check that 
\[
\cop({\bf x}) P = (-1)^{|{\bf x}|} P \cop({\bf x}),
\]
where $|{\bf x}|$ is the Hamming weight of the string ${\bf x}$.
Introduce Krauss operators
\[
K_p=i P \cop_p \quad \mbox{such that} \quad
K_p\, \cop_q =(-1)^{\delta_{pq}}\cop_q\, K_p.
\]
A one-mode quantum channel 
\[
\Phi_p(\rho)=\frac12 (1+b_p) \rho + \frac12 (1-b_p) K_p\, \rho K_p
\]
implements the transformation Eq.~(\ref{Phi}) with
$b_q=1$ for all $q\ne p$.
Note that $\Phi_p$ is a trace preserving completely
positive (TPCP) map iff $|b_p|\le 1$.
Since the one-mode channels commute,
$\Phi_p\circ \Phi_q=\Phi_q\circ\Phi_p$, we conclude that
\be\label{product_structure}
\Phi=\Phi_1\circ\Phi_2\circ\cdots\circ \Phi_{2n}.
\ee
Accordingly, $\Phi$ is a TPCP map for all $|b_p|\le 1$.
Expanding the product Eq.~(\ref{product_structure})
yields the desired Krauss representation of $\Phi$,
\be\label{Krauss}
\Phi(\rho)=\sum_{\bf x} p({\bf x}) \, \cop({\bf x}) \, \rho\,
\cop({\bf x})^\dag,
\ee
where
\[
p({\bf x})=\frac1{2^{2n}}\prod_{q=1}^{2n}(1+(-1)^{|{\bf x}| + x_q} \, b_q).
\]
The channels related by a transformation $b_p\to -b_p$ for some $p$
are unitarily equivalent, so we can focus on
non-negative values $0\le b_p\le 1$.

\subsection{Covariance}

The subsection summurazes some ideas proposed in
the papers~\cite{Holevo02,Cortese02}.
Let $\calS=\{U_1,U_2,\ldots,U_d\}$ be a set of $n$-qubit unitary operators
(which may or may not constitute a group) that completely
randomize any quantum state:
\be\label{randomize}
\frac1d \sum_{j=1}^d U_j \, \rho\,  U_j^\dag = I/2^n
\quad \mbox{for any state} \quad \rho.
\ee
A quantum channel $\Theta$ is said to be {\it covariant} if it commutes
with all operators $U\in \calS$, i.e.,
\[
\Theta(U A U^\dag)=U \Theta(A) U^\dag, \quad U\in \calS,
\]
for any operator $A$.
Covariance is a very useful property of a channel that allows one to 
reduce the problem of calculating the capacity to the problem
of finding the minimum output entropy of a channel, see Eq.~(\ref{Smin}).
Indeed, let  $|\Psi\ra$ be a state with the minimum output
entropy.
Consider an esemble of pure states $\{U|\Psi\ra\}_{U\in \calS}$,
where $U$ is chosen randomly and uniformly from the set $\calS$.
This ensemble has the maximally mixed average state since
$d^{-1}\sum_{j=1}^d U_j |\Psi\ra\la \Psi|U_j^\dag =2^{-n}\, I$.
Therefore the Holevo quantity of this ensemble is 
$n-S(\Theta(|\Psi\ra\la \Psi|))=n-S_{min}(\Theta)$.
Obviously, this is the absolute maximum of the Holevo quantity
for any ensemble of states in $Im(\Theta)$, so that
$C_1(\Theta) = n- S_{min}(\Theta)$
for any covariant channel.

The channel $\Phi$ defined in Eq.~(\ref{Phi}) is covariant.
Indeed, let $\calS$ be the set of $d=4^n$ operators $\cop({\bf x})$.
One can easily check that
\[
\frac1{4^n} \sum_{\bf x} \cop({\bf x})\, \cop({\bf y})\, 
\cop({\bf x}) = \left\{ \ba{rcl} 
0 &\mbox{if} & {\bf y} \ne 0, \\
I & \mbox{if} & {\bf y}=0.\\
\ea \right.
\]
It means that $\calS$ possesses the randomizing property
Eq.~(\ref{randomize}). On the other hand it is obvious that
\[
\Phi(\cop({\bf x})A\, \cop({\bf x})^\dag)=
\cop({\bf x})\Phi(A)\, \cop({\bf x})^\dag
\]
for any operator $A$. Therefore $\Phi$ is covariant channel,
and its one-shot capacity can be found from Eq.~(\ref{covariant}).

\subsection{Physical motivation}

Under certain circumstances
the fermionic product channels  may
adequately describe evolution  of fermi systems
interacting with environment.
As a toy
model consider 
a fermi system  that consists of
a single fermionic mode
($n=1$) which we 
describe by creation/annihilation operators $\aop^\dag$,
$\aop$. Suppose the system interacts with 
an environment which also can be described by a single
fermionic mode $\aop_e^\dag$, $\aop_e$.
Suppose the interaction between the system and the
environment is just the hopping Hamiltonian 
$H_{int}=g(\aop^\dag \aop_e + \aop_e^\dag \aop)$.
We assume that the initial state of the environment is maximaly
mixed, $\rho_e=(1/2)I$. The toy model channel is defined as
\[
\Phi(\rho) = \tr{}_e\left( U \rho\otimes \rho_e U^\dag \right), \quad
U=\exp{(-iH_{int}t)}.
\]
After straighforward calculations one gets
% \aop -> cos(gt) \aop + i sin(gt) \bop,
% \bop -> i sin(gt) \aop + cos(gt) \bop. 
\[
\Phi(I)=I,\quad 
\Phi(\aop)=\lambda \aop, 
\quad \Phi(\aop^\dag\aop-\aop\aop^\dag)=\lambda^2(\aop^\dag\aop-\aop\aop^\dag),
\]
where $\lambda=\cos{(gt)}$. Introducing $\cop$ generators
$\cop_1=\aop_1+\aop^\dag$, $i\cop_2=\aop_1-\aop_1^\dag$
we get: $\Phi(\cop_j)=\lambda\, \cop_j$ and $\Phi(\cop_1\cop_2)
=\lambda^2\, \cop_1\cop_2$. This is an example of the 
product fermionic channel Eq.~(\ref{Phi}) with $n=1$ and
$b_1=b_2=\lambda$. 
More generally, if the environment possesses a superconducting
order parameter, the interaction Hamiltonian may
include terms with particle-hole conversion, like in Andreev
reflection of electrons. In this case one can tune the interaction to
obtain an arbitrary string of coefficients $b_p$.

In fact, any 
unital channel that can be described by a quadratic
interaction Hamiltonian between the system and
the environment 
is unitarily equivalent to a channel from the family
Eq.~(\ref{Phi}), see~\cite{Bravyi05} for a proof.

\section{Gaussian states of fermions}
\label{sec:gauss}

%correlation matrix or covariance matrix?
In this section we describe a variational family of $n$-qubit
states that will be used to minimize  the output entropy
of the channel Eq.~(\ref{Phi}). These are Gaussian states.
The simplest example of a Gaussian state
is the Fock vacuum $|0^{\otimes n}\ra$ or 
any other product state diagonal in the standard basis. Denote
\be\label{standard_Gaussian}
\rho_\lambda= \frac1{2^n}\prod_{j=1}^n (I + \lambda_j \sigma^z_j)=
\frac1{2^n}\prod_{j=1}^n (I -i  \lambda_j \, \cop_{2j-1}\cop_{2j}).
\ee
An arbitrary Gaussian state can be converted into the
standard form $\rho_\lambda$ by a 
unitary evolution with a Hamiltonian quadratic 
in the operators $\cop_p$.
Here is a strict definition.
\begin{dfn}\label{dfn:Gaussian}
A state $\rho$  is Gaussian iff it can be
represented as
\be\label{general_Gaussian}
\rho =
U\, \rho_\lambda \, U^\dag, \quad
U=\exp{(iH_2 +i H_1)},
\ee
where 
$H_2$ and $H_1$ are Hermitian 
linear combinations of operators $\cop_p\cop_q$ and $\cop_p$
respectively.  
\end{dfn}
One can easily check that any pure Gaussian state has a form
$U\, |0^{\otimes n}\ra$, where $U$ is as above.

It will be more convenient to work with
even Gaussian states which correspond to choosing $H_1=0$
in Definition~\ref{dfn:Gaussian}.
\begin{dfn}\label{dfn:even}
A state $\rho$ of $n$ qubits is even Gaussian iff it can be
represented as
\be\label{even_Gaussian}
\rho =
U\, \rho_\lambda \, U^\dag, \quad
U=\exp{(iH_2)},
\ee
where 
$H_2$ is a Hermitian 
linear combination of $\cop_p\cop_q$.
\end{dfn}
One can easily check that 
any pure even Gaussian state has a form
$\exp{(iH_2)}\, |0^{\otimes n}\ra$. 
Our strategy will be to prove all statements first for
even Gaussian states.  Then we shall show how to establish
a correspondence between
Gaussian states of $n$-qubit system and even Gaussian
states of $(n+1)$-qubit system.

Unitary operators $U=\exp{(iH_2)}$ will be refered to 
as Bogolyubov tramsformations. 
Their conjugated action is 
\[
U\, \cop_p \, U^\dag = \sum_{q=1}^{2n} R_{pq}\, \cop_q,
\]
where $R$ is a rotation, $R\, R^T=I$, $\det{(R)}=1$. 
Any rotation $R\in SO(2n)$ can be realized by a proper Bogolyubov
transformation $U$.
In the rest of the section we list some basic properties
of Gaussian states.

\subsection{Wick's theorem}
As in the case of Gaussian probability distributions, an even Gaussian
state $\rho$ is completely characterized by its 
covariance matrix,
\[
M_{pq}=-\frac{i}2\, 
\tr\left[ \rho\, (\cop_p \cop_q -\cop_q \cop_p )\right].
\]

All higher moments of $\rho$ can be expressed in terms of $M$
using Wick's  theorem.
Namely, for any even binary string ${\bf x} \in \{0,1\}^{2n}$,
$|\xx|=2l$, one has
\be\label{Wick}
\tr(\rho\, \cop(\xx)) = i^{l} Pf(M[\xx]),
\ee
where $M[\xx]$ is a $2l\times 2l$ submatrix of $M$ obtained by selecting
all matrix elements $M_{pq}$ for which 
$x_p=x_q=1$, and $Pf$ stands
for the Pfaffian of a matrix. For example,
\[
i^{-2} \, \tr(\rho \, \cop_1\cop_2\cop_3\cop_4)=M_{12}M_{34}-M_{13}M_{24}
+ M_{14}M_{23}.
\]
If $\rho$ is an 
even Gaussian state then all odd correlators vanish,
\[
\tr(\rho\, \cop(\xx)) = 0 \quad \mbox{whenever} \quad
|\xx|=2l+1,
\]
which can be easily derived from the fact that
$\rho$ commutes
with the parity generator Eq.~(\ref{parity}), $P \rho=\rho P$.

We shall use Wick's theorem to prove that the channel
$\Phi$ defined in Eq.~(\ref{Phi}) maps the set of Gaussian states
into itself.
Also Wick's theorem will allow us to describe the action of $\Phi$ as 
a simple transformation of covariance matrices.

\subsection{Admissible covariance matrices}
Recall that our goal is to use Gaussian states 
as variational states to minimize the output entropy
of the channel $\Phi$. Since a Gaussian state is
completely characterized by its covariance matrix,
we have to find a set of admissible covariance matrices.
Note that if $M$ is chosen
arbitrarily, the operator $\rho$ defined by Wick's theorem
may have negative eigenvalues, i.e., it might not
describe a quantum state at all.
\begin{lemma}\label{lemma:admiss}
A real antisymmetric matrix $M$ is a covariance
matrix of an even Gaussian state iff
\be\label{MM<1}
M^T M\le I.
\ee
The corresponding state is pure iff $M^T M=I$.
\end{lemma}
For a proof of the lemma see~\cite{Bravyi05}.
Since $M$ is a real antisymmetric matrix, its
spectrum consists of $n$ pairs of conjugated eigenvalues
$\pm i\lambda_j$, $\lambda_j\ge 0$.
We shall refer to the numbers $\lambda_1,\ldots,\lambda_n$
as singular values of $M$.
The consistency condition Eq.~(\ref{MM<1})
is equivalent to inequalities $\lambda_j\le 1$.
The singular values $\lambda_j$ completely
determine the spectrum of an even Gaussian state
which is a product of $n$ binary spectrums
$(1/2)(1\pm \lambda_j)$, see~\cite{Bravyi05}.
Accordingly, 
the von Neumann entropy $S(\rho)=-\tr(\rho\log{\rho})$ 
is 
\be\label{entropy}
S(\rho)=\sum_{j=1}^n H\left(\frac{1+\lambda_j}2\right).
\ee

\subsection{Reduction to even states}
A simple correspondence between the sets of Gaussian states and even Gaussian
states can be established by adding one extra fermionic mode to
the system (this idea was proposed by Knill~\cite{Knill01}).
Let $\rho$ be an $n$-qubit Gaussian state.
Consider a linear map $\calE$ that maps 
$n$-qubit states into $(n+1)$-qubit states according to
\be\label{calE}
\calE(\rho)=V\, \rho\otimes (I/2)\, V^\dag, 
\quad 
V=\exp{\left(i\frac{\pi}4 \cop_{2n+1}\right)}.
\ee
Here $I/2$ stands for one maximally mixed qubit
which is labeled by $n+1$.
Since $V$ is a unitary operator, one has 
\be\label{S(calE)}
S(\calE(\rho))= S(\rho) +1 \quad \mbox{for any} \quad \rho.
\ee

Let us show that $\calE(\rho)$ is an even Gaussian state
for any Gaussian state $\rho$.
Indeed,
representing $\rho$ in the form Eq.~(\ref{general_Gaussian})
one gets
\[
\calE(\rho)=(VU)\, \rho_\lambda \otimes (I/2) \, (VU)^\dag,
\]
where $U=\exp{(iH_2+iH_1)}$.
Taking into account that $V$ commutes with $\rho_\lambda$,
we can rewrite it as
\be\label{aux1}
\calE(\rho)=(VUV^\dag) \, \rho_\lambda\otimes (I/2)\, (VUV^\dag)^\dag.
\ee
Recall that  $H_2$ is a linear
combination of $\cop_p\cop_q$ with $1\le p,q\le 2n$, while
$H_1$ is a linear combination of $\cop_1,\ldots,\cop_{2n}$.
It follows that $H_2$ commutes with $V$. On the other hand,
$V$ does not commute with $H_1$. Taking into account that
\[
V\, \cop_p \, V^\dag = i\cop_p\, \cop_{2n+1} 
\quad \mbox{for any} \quad p=1,\ldots,2n,
\]
one can easily check that $V\, H_1 V^\dag$ is a linear combination
of operators $\cop_p\, \cop_{2n+1}$, $p=1,\ldots,2n$.
It means that $V(H_2+H_1)V^\dag$ is a linear combination
of quadratic operators $\cop_p\, \cop_q$ only, 
where $1\le p,q,\le 2n+1$.
Comparing Eq.~(\ref{aux1}) and Definition~\ref{dfn:even}
we conclude that $\calE(\rho)$ is an even Gaussian state.

Conversly, if $\eta\equiv \calE(\rho)$ is an even Gaussian state
then $\rho$ is a Gaussian state. Indeed,
by definition of $\calE$, the generator $\cop_{2n+2}$
commutes with $\eta$. 
It means that a covariance
matrix $M$ of $\eta$ has all zeroes in the column $2n+2$
and the row $2n+2$, i.e., $M_{p,2n+2}=M_{2n+2,p}=0$.
Therefore it can be represented as
\[
M=R\, M_0 \, R^T, \quad M_0=\left( \ba{ccc} N & 0 & 0 \\
                                         0 & 0 & 0 \\
					 0 & 0 & 0 \\
\ea\right),
\]
where $R$ is a Bogolyubov transformation 
that involves only the generators $\cop_1,\ldots,\cop_{2n+1}$.
Accordingly, $\eta$ can be represented as
\[
\eta=U\, \eta_0\otimes (I/2)\, U^\dag, \quad U=\exp{(iH_2)},
\]
where $\eta_0$ is an even Gaussian state of
$n$ qubits with the covariance
matrix $N$, while $H_2$ is a Hermitian linear
combination of $\cop_p\, \cop_q$ with 
$1\le p,q\le 2n+1$. Repeating the same arguments as
above we get
\[
\rho\otimes (I/2)= W\, \eta_0\otimes (I/2)\, W^\dag,
\quad 
W=V^\dag \, U \, V.
\]
Conjugating $H_2$ by $V$ we do not affect terms
$\cop_p\, \cop_q$ that do not contain $\cop_{2n+1}$.
On the other hand, a term like $\cop_p\cop_{2n+1}$
is converted into $i\cop_p$. Thus the generator
$\cop_{2n+1}$ does not enter into $W$, and we get
$\rho = W\, \eta_0\, W$. It follows that
$\rho$ is a Gaussian state.

\subsection{Action of product channels on Gaussian states}

Let  $\rho$ be an even Gaussian state
with a covariance matrix $M$ and $\Phi$ be a product
fermionic channel with attenuation coefficients
$b_1,\ldots,b_{2n}$.
Moments of the state
$\rho$ are related to moments of the state $\Phi(\rho)$
by the following identity
\[
\tr(\Phi(\rho)\cop(\xx))=
\left( \prod_{p\, : \, x_p=1} b_p\right)  \tr(\rho \cop(\xx)).
\]
Comparing it with Wick's theorem Eq.~(\ref{Wick}) we conclude
that $\Phi(\rho)$ is an even Gaussian state with a
covariance matrix
\be\label{Phi(M)}
\Phi(M)=B \, M \, B^T, 
\ee
where $B$ is a diagonal matrix with entries $b_1,\ldots,b_{2n}$,
\[
B=\mbox{diag}(b_1,b_2,\ldots,b_{2n-1},b_{2n}).
\]

\begin{lemma}
Fermionic product channels map Gaussian states
into Gaussian states.
\end{lemma}
\begin{proof}
Let $\Phi$ be a channel as above.
Define a fermionic product channel $\hat{\Phi}$
acting on $n+1$ qubits such that
\begin{eqnarray}\label{hatPhi}
\hat{\Phi}(\cop_p) &=& b_p \, \cop_p, \quad p=1,\ldots,2n, \nn \\
\quad \hat{\Phi}(\cop_{2n+1}) &=& \cop_{2n+1}, \nn \\
\quad \hat{\Phi}(\cop_{2n+2}) &=& 0.
\end{eqnarray}
One can easily check that $\calE$ commutes with $\Phi$
in the following sense:
\be\label{EPhi=PhiE}
\calE\circ \Phi =\hat{\Phi}\circ \calE.
\ee
Let $\rho$ be a Gaussian state. Then $\calE(\rho)$
is an even Gaussian state, and thus 
$(\hat{\Phi}\circ\calE)(\rho)$ is an even Gaussian state.
It follows from Eq.~(\ref{EPhi=PhiE}) that
$(\calE\circ\Phi)(\rho))$ is also an even Gaussian state.
As we proved above, it implies that $\Phi(\rho)$ itself
is a Gaussian state.
\end{proof}

\subsection{Three-qubit states}
Suppose we restrict our attention only to even states,
i.e., those satisfying 
\[
P\, |\Psi\ra=|\Psi\ra,
\]
where $P$ is the parity generator, see Eq.~(\ref{parity}).
Obviously, for one qubit, $n=1$, the only even state is
$|0\ra$. This is a Gaussian state. If one takes two-qubits,
even states constitute a two-dimensional subspace
with a basis $|0,0\ra$ and $|1,1\ra$. 
Let us consider operators $A=(-i)\cop_1\cop_2$, $B=(-i)\cop_2\cop_3$,
and $C=(-i)AB=(-i)\cop_3\cop_1$. They all commute with $P$ and obey
the same commutation rules as the Pauli operators $\sigma^x$,
$\sigma^y$, $\sigma^z$ respectivly. Therefore any unitary operator
on the even subspace can be represented as
$U=\exp{(i\alpha A + i\beta B + i\gamma C)}$ for some 
real numbers $\alpha$, $\beta$, $\gamma$. Accordingly,
any even state has a form $U\, |0,0\ra$. It follows from
Definition~\ref{dfn:even} that any even two-qubit state
is Gaussian. Surprisingly enough, the same is true for three
qubits as well.
\begin{lemma}
Any even pure state of three qubits is Gaussian.
\end{lemma}
\begin{proof}
Let $\rho=|\Psi\ra\la \Psi|$ be an even pure Gaussian state
of three qubits.
Taking into account that all odd moments of $\rho$ vanish
and that $P\, |\Psi\ra=|\Psi\ra$, 
we can express $\rho$ as
\[
\rho=\frac18 (I+P)(I-\frac{i}2 \sum_{p,q} M_{pq} \, \cop_q\cop_q),
\]
where the sum runs over all $1\le p,q\le 6$ and $M$ is a 
real antisymmetric matrix. It is a well-known fact from
linear algebra that such a matrix $M$ can be represented as
\[
M=R \, \left( \ba{ccc} \lambda_1 \omega & 0 & 0 \\
                   0 & \lambda_2 \omega & 0 \\
                   0 & 0 & \lambda_3 \omega \\ \ea \right) \, R^T,
\quad \omega=\left(\ba{cc} 0 & -1 \\ 1 & 0 \\ \ea \right),
\]
where $\lambda_j$ are real numbers and $R\in SO(6)$ is 
a rotation. Consider a Bogolyubov transformation
$U$ such that
$
U\, \cop_p \, U^\dag = \sum_{q=1}^8 R_{pq}\, \cop_q.
$
Then
\[
\rho'=U\, \rho \, U^\dag =
\frac18 
(I+P)(I -i\lambda_1\, \cop_1\cop_2 - i\lambda_2\,\cop_3\cop_4
-i\lambda_3 \, \cop_5\cop_6).
\]
Here we used the fact that $UPU^\dag = \det{(R)} P =P$.
The state $\rho'$ is diagonal in the standard basis.
On the other hand $\rho'$ is a pure state.
Therefore $\rho'$ is a vector of the standard basis with
even number of ones. One can easily check that any such
vector is a Gaussian state. It follows that $\rho$ is
a Gaussian state as well.
\end{proof}
The minimum number of qubits supporting
even pure non-Gaussian states is $n=4$.
For example,  consider a state 
\[
|\theta\ra = \cos{(\theta)}\, |0,0,0,0\ra+
\sin{(\theta)}\, |1,1,1,1\ra.
\]
One can easily check that the covariance matrix of
$|\theta\ra$ has singular values smaller than one
whenever $\sin{(2\theta)}\ne 0$. It follows from
Lemma~\ref{lemma:admiss} that such states are
not Gaussian.

\section{Computation of the Gaussian capacity}
\label{sec:minimum}
Throught this section $\Phi$ is a fermionic product
channel with attenuation coefficients $b_1,\ldots,b_{2n}$
ordered in non-increasing way, i.e.,
\[
b_1\ge b_2\ge \ldots \ge b_{2n}.
\]
A computation of the Gaussian capacity will proceed in 
three steps. Firstly we shall compute the minimum 
output entropy achievable on even Gaussian input states.
Then we shall generalize it to arbitrary Gaussian input
using the trick with addition of a qubit.
Finally we shall verify that the optimal signaling ensemble
consists of Gaussian states.

\subsection{Even Gaussian input}
Our intermediate goal is to find the minimal value
of the output von Neumann entropy $S(\Phi(\rho))$ provided that
the input $\rho$ is an even Gaussian state.
Denote this minimal value
$S_{min,e}(\Phi)$.

Let $M$ be the covariance matrix of $\rho$. Since the
minimum output entropy is achieved on a pure input state,
we can assume that
\be\label{pure_input}
M^T\, M =I,
\ee
see Lemma~\ref{lemma:admiss}.
As we have shown in Section~\ref{sec:gauss}, the
output state $\Phi(\rho)$ is an even Gaussian state
with a covariance matrix
\be\label{final_M}
\Phi(M)=B\, M\, B^T, \quad B=\mbox{diag}(b_1,\ldots,b_{2n}).
\ee
Let $\lambda_1,\ldots,\lambda_n$ be singular values of
the matrix $\Phi(M)$ (recall that its eigenvalues are
$\pm i\lambda_j$). Taking into account Eq.~(\ref{entropy})
we get
\[
S_{min,e}=\min_{\lambda_1,\ldots,\lambda_n} \sum_{j=1}^n
H\left( \frac{1+\lambda_j}2\right),
\]
where the minimum is taken over all strings of singular
values $\lambda_1,\ldots,\lambda_n$ that are consistent
with Eqs.~(\ref{pure_input},\ref{final_M}).

We claim that
\be\label{smin_even}
S_{min,e}(\Phi)=
\sum_{j=1}^n H\left( \frac{1+b_{2j-1}b_{2j}}2 \right).
\ee
The proof will be based on convexity arguments and two
basic facts  from the majorization theory, see~\cite{MO79}.
Recall that
if $y=(y_1,\ldots,y_n)$ and $z=(z_1,\ldots,z_n)$ are
strings of real numbers,
the majorization relation $y\prec z$ is equivalent to
inequalities
\be\label{majorization}
\sum_{j=1}^k \ydown_j \le \sum_{j=1}^k \zdown_j,
\quad j=1,\ldots,n
\ee
with an equality for $k=n$. Here $\ydown_j$ 
and $\zdown_j$ are the 
$j$-largest elements of the strings
$y$ and $z$ respectively.

%\noindent
%{\bf Fact~1: (Rado theorem)} 
%If $y$ and $z$ are strings of real numbers such that
%$y\prec z$ then
%$y$ belongs to the convex hull of strings obtained by
%permuting the components of $z$.

\noindent
{\bf Fact~1:} If $f(x)$ is a concave function of a real
variable $x$ and $(y_1,\ldots,y_n)\prec (z_1,\ldots,z_n)$
then 
\[
\sum_{j=1}^n f(y_j) \ge \sum_{j=1}^n f(z_j).
\]

\noindent
{\bf Fact~2: (Horn-Visser-Zaanen theorem)} 

\noindent
For any square matrix $A$ define $D_l(A)$ as
a product of $l$ largest singular values of $A$.
Then  for any $m\times m$ matrices $A$ and $B$ 
\be\label{HVZ}
D_l(AB)\le D_l(A) \, D_l(B) \quad \mbox{for all}
\quad l=1,\ldots,m.
\ee
Note that $D_m(A)=|\det{(A)}|$, so Eq.~(\ref{HVZ})
becomes an equality for $l=m$.

Now let us proceed to a proof of Eq.~(\ref{smin_even}).
Let $\ldown_j$ be the sequence of singular
values $\lambda_j$ ordered in non-increasing way.
Taking into account that
\[
D_l(B)=D_l(B^T)=\prod_{j=1}^l b_j,
\quad 
D_l(M)=1,
\]
we can rewrite the Horn-Visser-Zaanen inequality 
\[
D_{2k}(B\, M \, B^T)\le D_{2k}(B)\, D_{2k}(M) \, D_{2k}(B^T)
\]
as
\be\label{aux2}
\prod_{j=1}^{k} {(\ldown_j)} \le
\prod_{p=1}^{2k} b_p,
\ee
with equality for $k=n$.
Let us introduce auxiliary variables
$y_j=\log{(\ldown_j)}$
and $z_j=\log{(b_{2j-1} b_{2j})}$
 taking values 
on the interval $(-\infty,0]$.
Obviously, $y_1\ge\ldots\ge y_n$ and
$z_1\ge\ldots\ge z_n$.
The inequalities Eq.~(\ref{aux2})
become equivalent to
the majorization inequalities Eq.~(\ref{majorization}),
so that 
\[
(y_1,\ldots,y_n)\prec (z_1,\ldots,z_n).
\]
The von Neumann entropy of the output state is
\[
S(\Phi(\rho))=\sum_{j=1}^n f(y_j),
\quad
f(x)=H\left( \frac{1-e^x}2 \right).
\]
To prove that
the function $f(x)$ is concave let us represent it as
$f(x)=H(g(x))$, where $g(x)=(1/2)(1-e^x)$.
Obviously, $g(x)$ is concave and maps the interval
$(-\infty,0]$ into the interval $[0,1/2)$.
The function $H(x)$ is concave and monotone increasing
on the interval $[0,1/2]$. Therefore $H(g(x))$ is
concave on the interval $(-\infty,0]$.
Applying Fact~1 we get 
\begin{eqnarray}\label{S_lower}
S(\Phi(\rho)) &=& \sum_{j=1}^n f(y_j) 
 \ge \sum_{j=1}^n f(z_j) \nn \\
&=&  \sum_{j=1}^n H\left( \frac{1+b_{2j-1}b_{2j}}2 \right).
\end{eqnarray}
This value of the output entropy can be achieved if the input $\rho$
is the Fock vacuum, i.e., 
\[
\rho=\frac1{2^n} \prod_{j=1}^n (I-i\cop_{2j-1}\cop_{2j})=
|0^{\otimes n}\ra\la 0^{\otimes n}|.
\]
Indeed, the corresponding output state is
\begin{eqnarray}
\Phi(\rho) &=& \frac1{2^n} \prod_{j=1}^n (I-
ib_{2j-1}b_{2j} \cop_{2j-1}\cop_{2j}) \nn \\
&=& \frac1{2^n} \prod_{j=1}^n (I+
b_{2j-1}b_{2j} \sigma^z_j). \\
\end{eqnarray}
The von Neumann entropy of $\Phi(\rho)$ matches the
lower bound Eq.~(\ref{S_lower}).

\subsection{Proof of Theorem~\ref{thm:main}}
Now we are ready to prove Theorem~\ref{thm:main}.
Firstly we shall find the minimum output entropy
achievable on a Gaussian input (not necesserily
even). Denote this quantity $S_{min,g}(\Phi)$. We claim that
\be\label{smin_odd}
S_{min,g}(\Phi)=
H\left( \frac{1+b_1}2 \right) +
\sum_{j=1}^{n-1} H\left( \frac{1+b_{2j}b_{2j+1}}2 \right).
\ee

Indeed, let $G_n$ and $G_n^e$ be sets of $n$-qubit Gaussian states
and even Gaussian states respectively, such that
\[
S_{min,g}=\min_{\rho \in G_n} S(\Phi(\rho)).
\]
We shall make use of the linear map $\calE$ defined in
Eq.~(\ref{calE}) that adds one qubit to the system.
Taking into account Eq.~(\ref{S(calE)}) we get
\[
S_{min,g}=\min_{\rho\in G_n} S((\calE\circ \Phi)(\rho))-1.
\]
Applying the commutation rule Eq.~(\ref{EPhi=PhiE}) we arrive to
\[
S_{min,g}=\min_{\rho\in G_n} S((\hat{\Phi}\circ \calE)(\rho))-1.
\]
As we have shown in Section~\ref{sec:gauss}.C,
$\calE(\rho)$ is an 
even Gaussian state, so that
\[
S_{min,g}\ge \min_{\eta\in G_{n+1}^e} S(\hat{\Phi}(\eta))-1.
\]
Since we already now how to compute the minimum over
even Gaussian states, see Eq.~(\ref{smin_even}), 
one gets
\[
S_{min,g}\ge 
\sum_{j=1}^{n+1} H\left( 
\frac{1+\hat{b}^\downarrow_{2j-1}\hat{b}^\downarrow_{2j}}2 \right)-1.
\]
Here $\hat{b}_1,\ldots,\hat{b}_{2n+2}$
are the attenuation coefficients
for the map $\hat{\Phi}$, see Eq.~(\ref{hatPhi}).
Obviosuly, the largest of them is 
$\hat{b}^\downarrow_1=\hat{b}_{2n+1}=1$,
while the smallest one is $\hat{b}^\downarrow_{2n+2}=
\hat{b}_{2n+2}=0$.
Therefore we arrive to
\be\label{Slower1}
S_{min,g}\ge 
H\left( \frac{1+b_1}2 \right) +
\sum_{j=1}^{n-1} H\left( 
\frac{1+b_{2j} \, b_{2j+1} }2 \right).
\ee
This lower bound is achieved on a state
\be\label{rhostar}
\rho_*=\frac1{2^n} (I+\cop_1)(I-i\cop_2\cop_3)
\cdots (I-i\cop_{2n-2}\cop_{2n-1}).
\ee
We have to verify that $\rho_*$ is indeed a Gaussian
state in the sense of Definition~\ref{dfn:Gaussian}.
Indeed, one can easily check that 
\[
\rho_*=(V_1 V_2)\, |0^{\otimes n}\ra\la 0^{\otimes n}| 
\, (V_1 V_2)^\dag,
\]
where $V_2$ is a Bogolyubov transformation that
shifts the operators $\cop_p$ cyclically according to
\[
(\cop_1,\cop_2,\cop_3,\ldots,\cop_{2n})\to
(-\cop_{2n},\cop_1,\cop_2,\ldots,\cop_{2n-1}),
\]
while $V_1=\exp{(-i\frac{\pi}4 \cop_{2n})}$. The
operator $V_1$ 
is chosen such that $V_1 (\cop_{2n}\cop_1) V_1^\dag =-i\cop_1$.
One remains to observe that
\[
\Phi(\rho_*)=
\frac1{2^n} (I+b_1 \cop_1)
\prod_{j=1}^{n-1}
(I-ib_{2j} b_{2j+1} \cop_{2j}\cop_{2j+1}).
\]
All factors in this product can be diagonalized
simultaniously, so the spectrum of $\Phi(\rho_*)$
is a product of $n$ binary spectrums
$(1/2)(1\pm b_1)$ and $(1/2)(1\pm b_{2j}b_{2j+1})$.
The corresponding entropy
$S(\Phi(\rho_*))$  matches the lower bound
Eq.~(\ref{Slower1}).

It is obvious that $C_1^g\le n-S_{min,g}$.
The optimal signaling ensemble that achieves 
this upper bound  consists of $2^n$ Gaussian states 
\[
\frac1{2^n} (I\pm\cop_1)(I\pm i\cop_2\cop_3)
\cdots (I\pm i\cop_{2n-2}\cop_{2n-1}).
\]
These states constitute an orthonormal basis
and each of them yields the output entropy $S_{min,g}$.

To complete the proof of Theorem~\ref{thm:main}
one remains to show that $C_1=C_1^g$ for $n=2$,
or, equivalently, that $S_{min}=S_{min,g}$ for
two-qubit channels.
Denote $D_n$ and $D_n^e$ sets of all $n$-qubits
states and even states respectively.
Repeating the same arguments as above one gets
\[
\min_{\rho\in D_2} S(\Phi(\rho))
\ge \min_{\eta\in D_3^e} S(\hat{\Phi}(\eta))-1.
\]
Obviously, the minimum in the righthand side
is achieved on a pure state $\eta$. But we have
already shown in Section~\ref{sec:gauss}.D
that any pure even three-qubit state 
is Gaussian. 
Therefore one can substitute the minimum
over $D_3^e$ by a minimum over $G_3^e$
and proceed as above.
We have proved Theorem~\ref{thm:main}.

\subsection{Minimal entropy and majorization}
In the previous subsection we have found a
Gaussian state $\rho_*$, see Eq.~(\ref{rhostar}),
 such that
$S(\Phi(\rho_*))\le S(\Phi(\rho))$ for any other
Gaussian state $\rho$. Since the von Neumann entropy 
measures amount of randomness contained in a state,
we can expect that $\Phi(\rho^*)$ is the least
randomized state in the image of $\Phi$ (as far as 
Gaussian inputs are concerned).
Here we shall put this statement into a more
strict form and prove the following interesting fact.

\noindent
{\bf Proposition~1:}
\be\label{rho*}
\Phi(\rho)\prec \Phi(\rho_*) \quad \mbox{for any
Gaussian $\rho$}.
\ee

Here the majorization relation 
between the density operators $\rho\prec \eta$
means that the spectrum of $\rho$ is majorized
by the spectrum of $\eta$. The majorization
relation $\rho\prec \eta$ is the quantative version
of the statement ``$\rho$ is more randomized than
$\eta$''. It is well known that $\rho\prec\eta$ iff
$\rho=\sum_\alpha p_\alpha U_\alpha \eta U_\alpha^\dag$
for some unitaries $U_\alpha$ and some probability
distribution $p_\alpha$. In particular
$\rho\prec \eta$ implies $S_\alpha(\rho)\ge S_\alpha(\eta)$,
where $S_\alpha(\rho)=\frac1{1-\alpha}\log{(\tr \rho^\alpha)}$
is the R\'enyi entropy.

Proposition~1 is a simple consequence
of the inequality Eq.~(\ref{aux2}) and the
following technical lemma.

\begin{lemma}\label{lemma:majorization}
Let $\alpha_1,\ldots,\alpha_n$ and
$\beta_1,\ldots,\beta_n$ be real numbers 
satisfying inequalities
\[
1\ge \alpha_1\ge \ldots \ge \alpha_n\ge 0,
\quad
1\ge \beta_1\ge \ldots \ge \beta_n\ge 0,
\]
Let $x=(x_1,\ldots,x_n)$ and $y=(y_1,\ldots,y_n)$
be random $n$-bit variables with probability
distributions
\[
P(x)=\frac1{2^n}\prod_{j=1}^n (1+(-1)^{x_j} \alpha_j),
\]
and
\[
Q(y)=\frac1{2^n}\prod_{j=1}^n (1+(-1)^{y_j} \beta_j).
\]
If for all $k=1,\ldots,n$ one has
\[
\prod_{j=1}^k \alpha_j \le \prod_{j=1}^k \beta_j
\]
with an equality for $k=n$ then $P$ is majorized by $Q$,
\[
P\prec Q.
\]
\end{lemma}

Let us first show how to derive Eq.~(\ref{rho*})
from the lemma.

\begin{proof}[Proof of the proposition]
It will be convenient to use a notation
$P_n(\alpha_1,\ldots,\alpha_n)$ for
a product of $n$ binary distributions 
$(1/2)(1\pm  \alpha_j)$.
 One suffices to prove Eq.~(\ref{rho*})
for pure states $\rho$. As we have shown
in the subsection~A, $\Phi(\rho)$ is
an even  Gaussian state whose covariance matrix
has singular values
$\ldown_1,\ldots,\ldown_n$ obeying the inequality
Eq.~(\ref{aux2}). As was mentioned in Section~\ref{sec:gauss}.B,
the spectrum of $\Phi(\rho)$ is a product of
$n$ binary spectrums $(1/2)(1\pm \ldown_j)$.
Applying Lemma~\ref{lemma:majorization} to
$\alpha_j=\ldown_j$ and $\beta_j=b_{2j-1}b_{2j}$
we conclude that
\be\label{aux3}
\Phi(\rho)\prec P_n(b_1b_2,\ldots,b_{2n-1}b_{2n}).
\ee

Now we use the correspondence between
Gaussian $n$-qubit states and even Gaussian
$(n+1)$-qubit states.
Let $\rho$ be an arbitrary Gaussian state.
Consider a linear map $\calE$ defined in Eq.~(\ref{calE})
and a channel $\hat{\Phi}$ defined in Eq.~(\ref{hatPhi}).
Recall that $\calE\circ \Phi=
\hat{\Phi}\circ \calE$. A state
\[
\eta=(\calE \circ \Phi)(\rho)=(\hat{\Phi}\circ \calE)(\rho)
\]
is even Gaussian.
Applying  Eq.~(\ref{aux3}) 
to the channel $\hat{\Phi}$ we get
\be\label{aux4}
Spec(\eta)\prec P_{n+1}(b_1,b_2 b_3,  \ldots,b_{2n-2} b_{2n-1},0).
\ee
Here $Spec(\eta)$ is the string of eigenvalues of $\eta$.
On the other hand, the spectrum of $\eta$ and
the spectrum of $\Phi(\rho)$ are related by
\[
Spec(\eta)=Spec(\Phi(\rho))\times \{\frac12,\frac12\},
\]
since $\calE$ just adds one maximally mixed qubit to
$\Phi(\rho)$ and then applies a unitary operator
that does not change the spectrum. Taking into 
account Eq.~(\ref{aux4}) we get
\begin{eqnarray}
Spec(\Phi(\rho))\times \{\frac12,\frac12\} \prec && \nn \\
P_{n+1}(b_1,b_2 b_3,  \ldots,b_{2n-2} b_{2n-1},0) && \nn \\
=P_n(b_1,b_2 b_3,  \ldots,b_{2n-2} b_{2n-1})\times \{\frac12,\frac12\}. &&
\end{eqnarray}
This majorization relation involves $2^{n+1}$ inequalities.
If one picks out only $2^n$ {\it even} inequalities (i.e. those
including a sum of two, four, six, e.t.c. largest eigenvalues)
one gets exactly the majorization relation
\[
Spec(\Phi(\rho))
\prec
P_n(b_1,b_2 b_3,  \ldots,b_{2n-2} b_{2n-1}).
\]
It is equivalent to the statement of the proposition.
\end{proof}

\begin{proof}[Proof of Lemma~\ref{lemma:majorization}]
We shall first prove the lemma for $n=2$ by brute force
method. Then we shall use standard majorization
theorems to extend the proof to all $n>2$.

If $n=2$ then the ordered probability distributions
$P=\{\pdown_1,\ldots,\pdown_4\}$ and 
$Q=\{\qdown_1,\ldots,\qdown_4\}$ are
\be\nn
\ba{rcl}
\pdown_1 &= &\frac14 (1+\alpha_1)(1+\alpha_2),\\
\pdown_2 &= &\frac14 (1+\alpha_1)(1-\alpha_2),\\
\pdown_3 &= &\frac14 (1-\alpha_1)(1+\alpha_2),\\
\pdown_4 &= &\frac14 (1-\alpha_1)(1-\alpha_2),\\
\ea
\quad
\ba{rcl}
\qdown_1 &= &\frac14 (1+\beta_1)(1+\beta_2),\\
\qdown_2 &= &\frac14 (1+\beta_1)(1-\beta_2),\\
\qdown_3 &= &\frac14 (1-\beta_1)(1+\beta_2),\\
\qdown_4 &= &\frac14 (1-\beta_1)(1-\beta_2).\\
\ea
\ee
The condition of the lemma for $n=2$ looks as
\[
\alpha_1\le \beta_1, \quad \alpha_1\alpha_2=\beta_1\beta_2.
\]
Our goal is to prove that $P\prec Q$.
After a simple algebra the first majorization
inequality $\pdown_1\le \qdown_1$ can be written as
\be\label{aux5}
\alpha_1+\frac{C}{\alpha_1} \le \beta_1 + \frac{C}{\beta_1},
\quad C\equiv \alpha_1\alpha_2=\beta_1\beta_2.
\ee
It is satisfied because a function $f(t)=t+C/t$ 
is monotone increasing on the interval $t\ge \sqrt{C}$
and because $\alpha_1,\beta_1\ge \sqrt{C}$.
The second majorization inequality $\pdown_1+\pdown_2\le
\qdown_1 + \qdown_2$ is equivalent to
$\alpha_1\le \beta_1$. The third majorization inequality
$\pdown_1+\pdown_2+\pdown_3 \le \qdown_1+\qdown_2+\qdown_3$
reduces to Eq.~(\ref{aux5})
that we have already proved. Thus $P \prec Q$.

To prove the lemma for arbitrary $n>2$ we shall need
two more basic facts from the majorization theory~\cite{MO79}:

\begin{dfn}\label{dfn:Ttransform}
Let $y=(y_1,\ldots,y_n)$ and $z=(z_1,\ldots,z_n)$ 
be strings of real numbers.
A string $y$ is a $T$-transform of string $z$ iff
there exist a real number $0\le \lambda \le 1$ and integers
$1\le a< b\le n$ such that
\[
y_c=z_c \quad \mbox{for all} \quad c\ne a,b,
\]
and
\[
\left[ \ba{c} y_a \\ y_b \\ \ea \right]=
\left[ \ba{cc} \lambda & 1-\lambda \\ 
               1-\lambda & \lambda \\ 
\ea \right] \cdot 
\left[ \ba{c} z_a \\ z_b \\ \ea \right].
\]
\end{dfn}

\noindent
{\bf Fact~3:} $y\prec z$ iff $z$ can be converted to $y$ by
a sequence of $T$-transforms.

\noindent
{\bf Fact~4:} $y\prec z$ iff $y_j=\sum_{k=1}^n P_{jk} \, z_k$
for some doubly stochastic matrix $P$.

Consider auxiliary variables $y_j=\log{(\alpha_j)}$
and $z_j=\log{(\beta_j)}$. The condition of the lemma
is equivalent to  $y\prec z$. Keeping in mind Fact~3, one
suffices to prove that $P\prec Q$ whenever
a string $y=(y_1,\ldots,y_n)$ is
a $T$-transform of a string
$z=(z_1,\ldots,z_n)$.
Let $a,b,\lambda$ be as in Definition~\ref{dfn:Ttransform}.
Since a $T$-transform is a doubly-stochastic matrix,
one has
\[
(y_a,y_b)\prec (z_a,z_b).
\]
In terms of the original variables $\alpha$ and $\beta$
it means that 
\[
\max{(\alpha_a,\alpha_b)}\le \max{(\beta_a,\beta_b)},
\quad 
\alpha_a\alpha_b=\beta_a\beta_b.
\]
Applying the $n=2$ proof given above
to the marginal 
probability distributions $P_{ab}$ and $Q_{ab}$
of the bits $a,b$ we conclude that $P_{ab}\prec Q_{ab}$.
Besides, for any bit $c\ne a,b$
the marginal distributions $P_c$ and $Q_c$ coincide.
Fact~3 implies that $Q_{ab}$ can be converted to $P_{ab}$
by a doubly stochastic matrix. Therefore $Q$ can be
converted to $P$ by a doubly stochastic matrix and
thus $P\prec Q$.
\end{proof}

\section{Numerical simulations}
\label{sec:numerics}

As we have seen before,
the smallest number of qubits for which the Gaussian analysis fails 
is $n=3$. 
Therefore,  we shall carry out numerical simulations
for fermionic product
channels 
acting on three and four qubits.
Our goal is to verify whether the minimum of the output
entropy is achieved on a Gaussian input state.
Below we describe an algorithm that in many cases allows one
to find the minimum output entropy with a high precision.

\subsection{Iterative minimization of the output entropy}
We shall describe an algorithm that generates a sequence
of states $|\psi_k\ra$, $k=0,1,2,\ldots$, such that
\begin{itemize}
\item The output entropies $S_k= S(\Phi(|\psi_{k+1}\ra\la \psi_{k+1}|))$
are non-increasing, i.e. $S_{k+1}\le S_k$.

\item The algorithm converges to an extremal point of
a functional $S(\Phi(|\psi\ra\la\psi|))$.
\end{itemize}

The algorithm exploits the following simple observation.
Consider a functional 
\[
h(\rho,\eta)=-\tr\left[ \Phi(\rho)\, \log \Phi(\eta)\right]
\]
that depends upon a pair of density operators $\rho$ and $\eta$.
Let us show that the global minimum of $h$ coincides
with  $S_{min}$. Indeed, let us use an identity
\be\label{rel_entropy}
h(\rho,\eta)= S(\Phi(\rho)) + S(\Phi(\rho)||\Phi(\eta)),
\ee
where $S(\omega||\sigma)=\tr[\omega(\log \omega -\log \sigma)]$
is the relative entropy.
Since the relative entropy is non-negative,
$S(\omega||\sigma)\ge 0$ for any $\omega,\sigma$,
and $S(\omega||\sigma)=0$ iff $\omega=\sigma$,
we conclude that
\be\label{Smin=h}
S_{min}(\Phi)=\min_{\rho,\eta} h(\rho,\eta).
\ee 
The key observation is that finding the minimum of $h(\rho,\eta)$
with respect to each {\it individual} variable 
is an exactly solvable problem. Indeed, if one fixes $\eta$ then
\begin{eqnarray}
\min_{\rho} h(\rho,\eta) &=& -\max_{\rho} 
\tr\left[ \rho\, \Phi^*(\log \Phi(\eta))\right] \nn \\
{} &=&
-\lambda_{max}\left(\Phi^*(\log \Phi(\eta) \right),
\end{eqnarray}
where $\Phi^*$ is the linear map conjugated to $\Phi$ (with respect to the
trace inner product) and $\lambda_{max}(X)$ is the maximum eigenvalue
of an operator $X$. The optimal state $\rho$ 
is obviously the highest eigenvector 
of  $\Phi^*(\log \Phi(\eta))$.
 On the other hand, if one
fixes $\rho$ and minimizes $h(\rho,\eta)$ over $\eta$,
the absolute minimum is achieved at $\rho=\eta$,
as immediately follows from Eq.~(\ref{rel_entropy}).

We shall attempt to find the minimum of $h(\rho,\eta)$
by carring out a sequence of alternating one-variable minimizations.
This iterative minimization procedure generates
a sequence of pure states $|\psi_0\ra,|\psi_1\ra,\ldots,$
such that (i) the initial state $|\psi_0\ra$ is 
chosen randomly; (ii) $|\psi_{k+1}\ra$ is the highest eigenvector
of an operator $\Phi^*(\log \Phi(|\psi_k\ra\la \psi_k|))$
(if the highest eigenvalue is degenerate, $|\psi_{k+1}\ra$
is chosen randomly from the corresponding eigenspace).
Consider a sequence
\[
S_k=S(\Phi(|\psi_k\ra\la\psi_k|)), \quad k=1,2,\ldots,\infty.
\]
Since each one-variable minimization does not increase
the objective function, we have $S_{k+1}\le S_k$, i.e.,
the sequence $\{S_k\}$ is monotone decreasing.

Denote $S_*=\lim_{k\to \infty} S_k$.
Let $|\psi_*\ra$ be any limiting point of the sequence
$\{|\psi_k\ra\}$ and $\rho_*=|\psi_*\ra\la \psi_*|$.
Obviously
\[
h(\rho_*,\rho_*)\le h(\rho,\rho_*), \quad \mbox{and}
\quad 
h(\rho_*,\rho_*)\le h(\rho_*,\eta)
\]
for any states $\rho$ and $\eta$. Therefore
a point $\rho=\eta=\rho_*$ is an extremum of
$h(\rho,\eta)$. Thus $\rho_*$ is an extremal
point of the output entropy $S(\Phi(\rho))=h(\rho,\rho)$. 
Obviously, $\rho_*$ can not be a (local) maximum of
$h$. However, it might be a local minimum or a saddle
point. In principle, saddle points can be eliminated
by computing a Hessian of $h(\rho,\rho)$ at the 
point $\rho_*$.

It seems very unlikely that the iterations can be
trapped in a very narrow and shallow local minimum, since 
one-variable minimization allows one to `tunnel' through
potential barriers. We can also expect that 
the method should work well for channels
that are sufficiently noisy, since a landscape of
the objective function $S(\Phi(|\psi\ra\la\psi|))$
becomes more flat as $\Phi$ becomes less noisy.
Symmetry of a channel also might be an important issue
because one-variable iterative minimization typically works well
for highly symmetric functions.

Before applying the algorithm to fermionic product
channels we have tested it 
on products of one-qubit unital channels for which
the exact value of $S_{min}$ is known, see~(?).
The algorithm always converges to the correct
value of $S_{min}$ for channels with sufficiently
large amount of noise. For example,
let $\Theta$ be a one-qubit Pauli depolarizing
channel, such that errors $\sigma^x$, $\sigma^y$,
and $\sigma^z$ occur with a probability $p$ each.
Then the iterative algorithm correctly computes
$S_{min}(\Theta^{\otimes 4})=4 H(1-2p)$
in a region $p\ge 0.05$.

\subsection{Numerical simulations}
In the rest of this section we discuss results of numerical simulations
performed for several families of 
fermionic product channels on three and four qubits.
For each particular channel
the minimum output entropy has been
evaluated using the iterative minimization algorithm
with $64$ iterations and $16$ independent choices of the initial
state. 

A plot on Figure~1 consists of two groups of points
marked by `+' and `$\times$'. Each of these point corresponds
to some particular  $3$-qubit channel from the family Eq.~(\ref{Phi}).
The points `+' correspond to a symmetric channel
\be\label{+}
\Phi_{+}(\cop_p)=b\, \cop_p, \quad p=1,\ldots,6,
\ee
where $0\le b\le 1$ is a parameter specifying the channel.
The points `$\times$' correspond to a channel
with a non-trivial distribution of the coefficients
$b_p$, namely
\be\label{x}
\Phi_{\times}(\cop_p)=b^{\, p/3}, \quad p=1,\ldots,6.
\ee
The parameter $b$ is plotted on the horizontal axis. 
On the vertical axis we plotted the smallest 
output entropy $S_{min}$ found by the algorithm.
The dotted lines represent 
the minimum output entropy that can be achieved on a Gaussian 
input state, i.e., the function $S_{min,g}$
given by Eq.~(\ref{smin_odd}). The inset plot
shows the deviation $S_{min}-S_{min,g}$.

\begin{figure}
\includegraphics[scale=0.35,angle=-90]{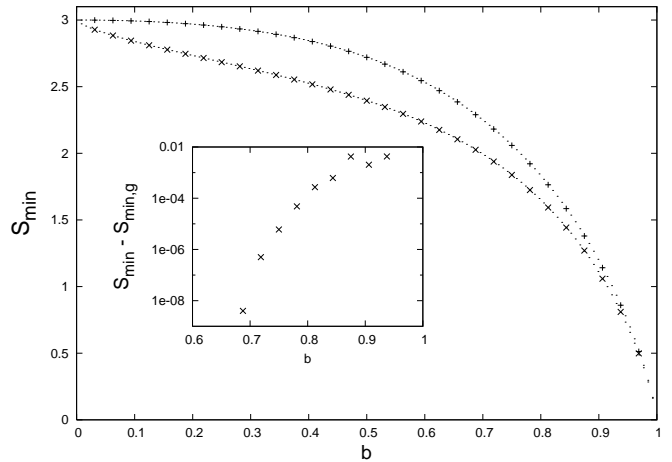}
\caption{Three-qubit channels.
Vertical axis: the smallest output entropy $S_{min}$
found by the iterative minimization algorithm
for the channels $\Phi_{+}$ and $\Phi_{\times}$.
Horizontal axis: the parameter $b$ specifying
the channels, see Eqs.~(\ref{+},\ref{x}).
Dotted lines: the minimum output entropy 
achieved on a Gaussian input ($S_{min,g}$).
On the inset plot: deviation $S_{min}-S_{min,g}$
for the channel $\Phi_{\times}$.
}
\end{figure}

It turns out that the algorithm always converges to
the same value of $S_{min}$ for the symmetric channel
$\Phi_+$.
A typical deviation $S_{min}-S_{min,g}$ for this
channel is about $10^{-9}$ which 
has the same order of magnitude as the numerical noise.

As for the non-symmetric channel $\Phi_{\times}$, the algorithm
always converges to the same value of $S_{min}$ unless
$b$ is close to $1$. A typical deviation $S_{min}-S_{min,g}$
is about $10^{-9}$ in the region $0\le b\le 0.7$,
so it is not shown on the plot. The deviation is always
positive however.
On the other hand, if $b$ is close to $1$, 
the iterations sometimes converge to a local minimum with 
a value of $S_{min}$ exceeding $S_{min,g}$ by several percents.

Numerical simulations have been also carried out for 
analogous $4$-qubit channels. A plot on Figure~2 
shows the smallest output entropy found by the algorithm
for a symmetric channel $\Phi_{+}$ defined as
\be\label{4+}
\Phi_{+}(\cop_p)=b\, \cop_p, \quad p=1,\ldots,8,
\ee
and for a channel $\Phi_{\times}$ 
with a non-trivial distribution of the coefficients
$b_p$, namely
\be\label{4x}
\Phi_{\times}(\cop_p)=b^{\, p/4}, \quad p=1,\ldots,8.
\ee
Here $0\le b\le 1$ is a parameter specifying the channel.

The algorithm always converges to
the same value of $S_{min}$ for the symmetric channel
$\Phi_+$.
A typical deviation $S_{min}-S_{min,g}$ for this
channel is about $10^{-9}$.

For the channel $\Phi_{\times}$ the algorithm 
always converges to the same value of $S_{min}$
in a region $b\le 0.55$. The deviation
$S_{min}-S_{min,g}$ drops down to $10^{-9}$ in
a region $b\le 0.6$, so it is not shown on the plot.

\begin{figure}
\includegraphics[scale=0.35,angle=-90]{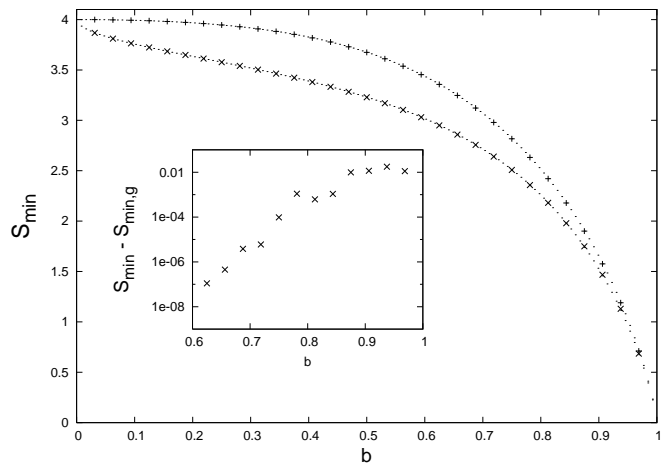}
\caption{Four-qubit channels.
Vertical axis: the smallest output entropy $S_{min}$
found by the iterative minimization algorithm
for the channels $\Phi_{+}$ and $\Phi_{\times}$.
Horizontal axis: the parameter $b$ specifying
the channels, see Eqs.~(\ref{4+},\ref{4x}).
Dotted lines: the minimum output entropy 
achieved on a Gaussian input ($S_{min,g}$).
On the inset plot: deviation $S_{min}-S_{min,g}$
for the channel $\Phi_{\times}$.
}
\end{figure}

The numerical results seem to support the conjecture that
the minimum output entropy is achieved on a Gaussian 
input state, i.e., that $S_{min}(\Phi)=S_{min,g}(\Phi)$
for all fermionic product channels.
The fact that the algorithm always converges to the same value
of $S_{min}$ for the symmetric channels $\Phi_{+}$ 
suggests that a landscape of the corresponding objective
function $S(\Phi_{+}(|\psi\ra\la\psi|))$ is particularly
simple. It might be an indication that $S_{min}$ can be
computed analytically for the symmetric channels.

\section{Acknowledgments}

Discussions with Christopher King
and Frank Verstraete are gratefully acknowledged.
This work was supported by the National Science Foundation under 
grant number EIA-0086038.

\end{document}